# ACTIVE CRYOVOLCANISM ON EUROPA?

Sparks, W.B.[1], Schmidt, B.E.[2], McGrath, M.A.[3], Hand, K.P.[4], Spencer, J.R.[5], Cracraft, M.[1], Deustua, S.E[1]

[1] Space Telescope Science Institute, 3700 San Martin Drive, Baltimore, MD 21218, USA.

[2] Georgia Institute of Technology, School of Earth and Atmospheric Sciences, 311 Ferst Drive, Atlanta, GA 30332, USA

[3] SETI Institute, 189 N Bernardo Ave, Mountain View, CA 94043, USA

[4] Jet Propulsion Laboratory, California Institute of Technology, 4800 Oak Grove Drive, Pasadena, CA 91109, USA

[5] Southwest Research Institute, 1050 Walnut St. Suite 300, Boulder CO 80302, USA

Short title: CRYOVOLCANISM ON EUROPA

Corresponding author: sparks@stsci.edu

ABSTRACT

Evidence for plumes of water on Europa has previously been found using the Hubble Space Telescope (HST) using two different observing techniques. Roth et al. (2014) found line emission from the dissociation products of water. Sparks et al. (2016) found evidence for off-limb continuum absorption as Europa transited Jupiter. Here, we present a new transit observation of Europa that shows a second event at the same location as a previous plume candidate from Sparks et al. (2016), raising the possibility of a consistently active source of erupting material on Europa. This conclusion is bolstered by comparison with a nighttime thermal image from the Galileo Photopolarimeter-Radiometer (PPR) which shows a thermal anomaly at the same location, within the uncertainties (Spencer et al. 1999). The anomaly has the highest observed brightness temperature on the Europa nightside. If heat flow from a subsurface liquid water reservoir causes the thermal anomaly, its depth is ≈1.8–2 km, under simple modeling assumptions, consistent with scenarios in which a liquid water reservoir has formed within a thick ice shell. Models that favor thin regions within the ice shell that connect directly to the ocean, however, cannot be excluded, nor modifications to surface thermal inertia by subsurface activity. Alternatively, vapor deposition surrounding an active vent could increase the thermal inertia of the surface and cause the thermal anomaly. This candidate plume region may offer a promising location for an initial characterization of Europa's internal water and ice and for seeking evidence of Europa's habitability.

KEYWORDS: Europa, cryovolcanism

1. INTRODUCTION

Europa is perhaps the most exciting astrobiological target in the Solar System. While Mars may have hosted liquid water oceans in the geological past, there is strong evidence that Europa has a saline, liquid water ocean at the present time (Kivelson et al. 2000). With a rocky mantle submerged beneath 100 km of water, and surface ice heavily processed by Jovian radiation, the energy and ingredients required for life may be present in abundance (Chyba 2000; Chyba & Hand 2001; Hand et al. 2007; Hand et al. 2009). There is substantial evidence that exchange occurs between the ocean and surface through a variety of mechanisms on geologically short timescales. Complex areas of brownish-red material may have been sourced from within the moon, arising from geologic processes that involve melting of the ice shell (Pappalardo et al. 1999; Figuerdo et al. 2003; Figuerdo & Greeley 2004; Schmidt et al. 2011; Kattenhorn & Prockter 2014). Active cryovolcanism has been considered as a transport mechanism, which would result in recent deposition onto the surface from liquid reservoirs below (Figuerdo & Greeley 2004; Fagents et al. 2000; Fagents 2003; Miyamoto et al. 2005). Although "viable", no unequivocal evidence of extrusive cryovolcanism has been detected (Fagents 2003). Cryovolcanos (or cryogeysers) would provide relatively easy access to such Europan water, without the need to penetrate the thick, global ice shell. It has also been hypothesized that shallow water reservoirs encased within the ice are related to the formation of chaos terrain, and could lead to localized water vapor venting (Collins & Nimmo 2009; Schmidt et al. 2011; Walker & Schmidt 2015). Transient venting may also arise from rapid boiling of cryolavas erupted onto the zero-pressure environment of the Europa surface (Allison & Clifford 1987; Fagents 2003; Quick et al. 2017).

The discovery of hydrogen and oxygen line emission beyond the limb of Europa (Roth et al. 2014) raised the likelihood of observable, active water vapor vents on Europa. Sparks et al. (2016) sought plumes in ten far-ultraviolet (FUV) images of Europa, obtained using the Space Telescope Imaging Spectrograph (STIS) on board HST as Europa transited Jupiter, by looking for absorption of Jupiter's smooth reflected sunlight. Statistically significant evidence for off-limb absorption was found on three occasions, adding to the evidence for contemporary plume activity. Of paramount importance in establishing the reality of a transient or intermittent phenomenon, is whether it repeats. A new FUV transit image of Europa, obtained on February 22, 2016, presented here, shows a dark patch of similar character and location to the one seen on March 17, 2014. We also show that this position coincides with a thermal anomaly identified using the Galileo spacecraft PPR instrument (Spencer et al. 1999).

2. OBSERVATIONS

Subsequent to the analysis of Sparks et al. (2016), two more Jovian transit observations were obtained of Europa with HST/STIS, using the F25QTZ far-ultraviolet long-pass filter.

The F25QTZ filter has a pivot wavelength of 159.6 nm, and full width half maximum (FWHM) of 23.2 nm, but with an uncertain amount of redleak, see discussion in Sparks et al. (2016). This filter has a somewhat redder short-wavelength cutoff than the F25SRF2 filter, pivot wavelength 145.2 nm and FWHM 28.2 nm, used previously, but is otherwise very similar. (The F25QTZ eliminates auroral and geocoronal [OI] and had the potential to reduce the number of STIS buffer dumps). The observations were obtained in TIME-TAG mode as the image of Europa was drifted across the STIS detector. Table 1 summarizes the observing parameters of the new data. Data processing followed the procedures of Sparks et al. (2016) for geometric correction, flat-fielding and techniques for assembling images from TIME-TAG data, including both the target Europa image and model images of the Jovian background. The February 22, 2016 observation was obtained with the image of Europa located in a different region of the STIS detector, and the orientation of the telescope was rotated 166 degrees relative to the March 17, 2014 observation.

In the far ultraviolet:

- Jupiter is smooth, dominated in appearance by scattering from high level hazes in the Jovian atmosphere.
- The motion of Jupiter relative to Europa during an observation further reduces its apparent contrast due to smear.
- The ice of Europa is dark.
- The scattering cross-sections of molecules of interest – $H_2O$, $O_2$ – are high, maximizing their capacity to attenuate background light.
- Micron sized ice grains, which contribute to the composition of Enceladus's plumes, scatter light effectively in the far ultraviolet.
- The spatial resolution of HST is near its highest. The diffraction limit of HST at 150

nm is ~13 millarcseconds (mas). At a distance of 4.5 AU, 13 mas corresponds to 42 km. Diffraction limited structure is seen even though the HST point spread function (PSF) in the FUV is complex (Sparks et al. 2016).

3. RESULTS

The March 17, 2014 image shows a compact, dark patch at latitude 16.5°S, protruding from the trailing limb to an estimated height ≈50 km, Fig. 1(a). The corresponding water column density was estimated to be $1.8 \times 10^{21} m^{-2}$, with a total of $1.8 \times 10^{32}$ H$_2$O molecules or ≈$5.4 \times 10^6$ kg (Sparks et al. 2016). Fig. 1(b) shows the image from February 22, 2016 smoothed and divided by a model of Jupiter obtained from the data. Fig. 1(d) shows the same image as Fig. 1(b) at full spatial resolution to illustrate both that the patch covers multiple 35 km pixels, and that the "noise" in the image is well-behaved. After division by the model of the Jovian background, the mean brightness in the patch is 0.80, i.e. 20% absorption relative to the surroundings. For an empirically measured noise level, this corresponds to a 4.9 σ decrease in brightness. Fig 1(c) shows the average of the two (2014, 2016) candidate plume images divided by the mean of the other ten transit images, implying the feature is not always visible. The new image of March 4, 2016, did not show any evidence for the presence of plumes and was included in the comparison set.

The latitude of the March 17, 2014 patch was 16.5°S, and that of the February 22, 2016 is 16.8°S, an insignificant difference of 0.3 pixels. The February 2016 patch is centered ≈2.8 pixels ≈99 km above the limb. Fig. 2(a) shows (green) the region containing the March 17, 2014 candidate (taken directly from Fig. 23 in Sparks et al. 2016), with the addition of a second ellipse (cyan) representing the uncertainty in location of the February 22, 2016 candidate, superimposed on a high resolution Galileo/Voyager USGS map. Fig. 2(b) shows nighttime brightness temperature contours obtained with the Galileo PPR covering the same region, from Spencer et al. (1999). There is a thermal anomaly with elevated brightness temperature north and west of the crater Pwyll. This location in fact has the highest brightness temperature in the observed nighttime, even though it is near local dawn and has hence been in darkness for a lengthy period of time. The position of the thermal anomaly is essentially identical, within the uncertainties, to that of the candidate plumes, and hence we discuss the implications of a potential causal relationship between the two.

4. DISCUSSION

Proposed mechanisms to create plumes include explosive release of volatile compounds dissolved within water as confining pressure is removed (Crawford & Stevenson 1988; Pappalardo et al. 1999; Fagents et al. 2000), and pressurization and fracturing due to expansion of ice as an enclosed water cavity freezes (Pappalardo et al. 1999; Fagents et al. 2000; Schmidt et al. 2011; Walker & Schmidt 2015). Subsequent collapse of such a cavity may be implicated in chaos formation (Schmidt et al. 2011). Vapor release, or surface venting, due to boiling associated with the effusion of liquid cryolavas can also, in principle, result in plume activity (Allison & Clifford 1987; Fagents 2003; Quick et al. 2017).

Inspection of Fig. 2 reveals a wide variety of terrains within the region of interest. In particular: (1) two dark lineae, one of which appears to be a continuation of Phineus Linea, running approximately east-west are presumably fractures like other Europan lineae; a low-density linear plume arising from these fractures could have high projected column density when seen almost end-on in the HST viewing geometry; (2) a number of small dark patches likely to be chaos regions. The uncertainty ellipse also contains several small dark spots that are probably examples of the "pits, spots, and domes" seen in high-resolution images of other parts of Europa (e.g. Collins & Nimmo 2009). These features, particularly when associated with low albedos, are thought to be localized disruptions in the ice crust, similar to, but smaller than, classical chaos regions, and tend to be geologically recent, and thus are also plausible sources of plume activity. All of the above features can potentially be associated with cryovolcanism of various forms (Pappalardo et al. 1999; Fagents et al. 2000; Fagents 2003; Quick et al. 2017) and hence be related to plume phenomena.

Spencer et al. (1999) considered both endogenic heat conducted through the lithosphere, and modification of the surface thermal inertia (which is sensitive to porosity, and the nature of grain-to-grain contacts), as possible causes for thermal anomalies and large-scale temperature variations across the Europa nightside. They concluded that on a global scale, variations in thermal inertia are the most likely cause of the nighttime temperature variations seen on Europa, a result supported by the more detailed analyses of Rathbun et al. (2010).

Locally, however, endogenic heat sources remain a possibility. The Galileo PPR data, Fig. 2(b), has a spatial resolution of 110 km pixel$^{-1}$ (Rathbun et al. 2010), hence the anomaly of Fig. 2(b) is extended over several hundred kilometers. Inspection of Fig. 1 of Spencer et al. (1999) indicates a base temperature of $T_1 \approx 93.5$K, reaching $T_2 \approx 95$K at the center of the "warm" region, with an average diurnal temperature of $T_a = 100-105$K. If endogenically generated, the observed temperature difference requires a heat flux $Q \approx 0.28$ Wm$^{-2}$. From the heat conduction equation (Fourier's Law), $Q = -\kappa \nabla T$, where $\kappa$ is the thermal conductivity, the consequent depth to liquid water at $T_b = 260-273$K, is $H \approx 1.8-2.0$ km, assuming that for ice $\kappa = 488.2/T + 0.47$ Wm$^{-1}$K$^{-1}$ (Hobbs 1974). The range in $H$ corresponds to the stated ranges of upper and lower temperature boundary conditions, $T_a$, $T_b$. An uncertainty of 0.5 K in each of $T_1$, $T_2$, the thermal anomaly amplitude, results in a 28% uncertainty in $Q$ and hence an uncertainty of 0.6 km in the ice thickness, $H$, in addition to the systematic range arising from the temperature boundary conditions above, under the endogenic heat assumption.

Tidal heating models exclude such a large heat flow globally, although there are plausible mechanisms that could produce a locally high heat flux. These include a near-surface liquid water reservoir (Schmidt et al. 2011), local convergence of tidal forces (Rhoden et al., 2015), and oceanic hydrothermal plumes above which locally thin ice may form (Goodman et al. 2004). Liquid reservoirs within the ice layer may arise from melting above rising ice plumes or "diapirs", which likely have both compositional and thermal buoyancy (Pappalardo & Barr 2004), and other manifestations of convection within the ice shell, or tidally-enhanced melting of such plumes. These are most likely to form between 3-5 kilometers from the surface where reasonable eutectic melting points exist (e.g. Sotin et al. 2002; Collins & Nimmo 2009; Zolotov & Kargel 2009; Schmidt et al. 2011). While conjectured as an option to create a melt reservoir within the ice, independently of the presence of plumes, such subsurface activity can, in and of itself, lead to modification of the surface thermal inertia (Fagents et al. 2000), hence also yielding a potential association of plumes and thermal characteristics. Injection of water into the subsurface may also be possible (e.g. Ojakangas & Stevenson 1989; Craft et al. 2016), although fractures through the ice shell are difficult to form (Goldreich & Mitchell 2009; Walker et al. 2012; Walker & Schmidt 2015). Models that favor thin regions within the ice shell that connect directly to the ocean also cannot be excluded (Carr et al. 1998; Greenberg et al. 2002).

Alternatively, the thermal anomaly may be caused by a locally higher value for the surface thermal inertia related to a deposition blanket from an erupting vent. To reach a height of 99 km, an ejection speed of $\approx$510 m s$^{-1}$ is required for a ballistic trajectory, while Roth et al. (2014) inferred a plume ejection speed of 700 m s$^{-1}$. Such speeds and consequent plume heights are higher than generally expected from theory (Fagents 2003; Quick et al. 2013), and imply a high vapor content for the plume (e.g. in the explosive venting of volatile bearing liquid water studied in detail by Fagents et al. 2000, where high gas content is also required). The thermal velocity of water vapor is approximately 500 m s$^{-1}$ at ~175K, characteristic of Enceladus plume velocities (Hansen et al. 2006; Nimmo et al. 2007) and similar to velocities implied by the current observations, though detailed plume models are beyond the scope of this work.

Although the escape velocity of Europa is ≈2.0 km s$^{-1}$, causing most of the vented material to fall back to the surface, the great distances to which H$_2$O molecules would nevertheless travel may result in very widely dispersed deposits and a negligible visible signature. Active cryovolcanism in the ballistic regime would produce a deposition layer extending to a radius approximately twice the height (Fagents et al. 2000; Quick et al. 2013), comparable in extent to the thermal anomaly. Freshly deposited material, such as condensed water vapor, extended over such an area, is likely to exhibit different thermal characteristics from its surroundings. While porous "snow-like" material is expected to have lower thermal inertia, contrary to inference, condensing water vapor from a plume could plausibly contribute to surface sintering, increasing the thermal inertia of the deposition region around the plume and thereby causing the thermal anomaly. There are no obvious anomalous albedo features, however that does not preclude a modification to the thermal inertia (by analogy, the dramatic thermal inertia anomaly seen on Mimas' leading side [Howett et al. 2010] is barely detectable in visible wavelengths). In this scenario, there are no required constraints on the ice thickness. A higher thermal inertia could also result from compositional differences, e.g. salts or other endogenous materials, but such hypotheses are hard to constrain given limited spectroscopic data (e.g., Carlson et al. 2009).

If the plumes and thermal anomaly are causally related, an activity timescale of at least ~20 years is implied (September 1998 to February 2016), an individual eruption timescale ~1 hour (45 min observation duration), and a plume duty cycle of 17±12% (two events from twelve observations).

There are significant caveats concerning the presence of the plume candidates and their association with thermal structure on Europa: there is no obvious evidence of unusual morphological or albedo signatures associated with the thermal anomaly, nor centered on the candidate plume, consistent with the null result of Phillips et al. (2000) who sought morphological evidence of ongoing geologic activity on Europa. It could, however, be that the plume deposits are sufficiently tenuous that they were undetectable by the combination of Voyager, Galileo and New Horizons (Phillips et al. 2000; Phillips 2015). Also, the statistical significance of the plume candidates is not completely definitive: while random Poisson fluctuations are unlikely given the coincidence of two $\gtrsim 4\sigma$ events discussed herein, the further possibility of subtle, transient systematics remains, given the complex PSF of HST in the FUV (Sparks et al. 2016). It is also possible that the elevated brightness temperature in the region of interest is completely unrelated to the plume phenomenon, as, presumably, are temperature variations elsewhere on Europa, in which case the positional coincidence would be just that: a coincidence.

To mitigate potential problems, and eliminate small scale detector artifacts, the image of Europa was drifted across the detector during the observation. The 2016 observation was also taken with Europa located in the opposite quadrant of the STIS detector, with the orientation of the telescope rotated 166 degrees relative to the 2014 observation. This disfavors, though does not completely eliminate, the possibility of a PSF anomaly coupled to a hypothetical surface feature influencing the exterior limb region. Face-on images of the trailing hemisphere, however, show no apparent dark spot, and for this observation, the solar illumination angle is such that the terminator is on the leading hemisphere, precluding a dark spot introduced by the solar illumination pattern on the trailing hemisphere. We do not know of any artifacts that would introduce the appearance of the off-limb dark patch (plume candidate) in the HST data, but stress that we are working at challenging wavelengths for HST and that there may be lessons yet to be learned. The combination of two very similar plume candidates in the same location, and the presence of a thermal anomaly at the same location, leads us to conclude that the evidence does, on balance, favor the plume phenomenon hypothesis, while acknowledging that unlikely events and coincidences do occur and that the statistical significance of the plume candidate ($\approx 4.9\ \sigma$) is not a secure $\gg 5\ \sigma$ result, which would be preferred.

5. CONCLUSIONS

We have presented additional evidence for sporadic or intermittent contemporary plume activity on Europa. A new HST plume candidate, seen as a possible absorption feature against the smooth background of Jovian scattered ultraviolet light, coincides in position not only with a similar previously identified candidate, but also with the peak brightness temperature position of a nighttime thermal anomaly identified by Spencer et al. (1999). The coincidence of two plume candidates at the same relatively well-defined location, separated by two years, improves the statistical likelihood that they represent a physical phenomenon associated with Europa, although we are working with data that is close to the limit of HST's capabilities and there may yet prove to be unsuspected image artifacts. With a priori reason to anticipate a thermal signature of plume activity, as on Enceladus, the spatial coincidence of the plume candidates with a zone of elevated temperature seen by Galileo on the Europa nightside adds to the intriguing nature of this location. Under the two assumptions that (a) the plume candidates are real and (b) the association with the elevated temperature is causal, the physical implications are that either the thickness of ice above liquid water is relatively small, or that vapor deposition is responsible for modifying the surface thermal inertia (without implications for ice thickness).

If borne out with future observations, these indications of an active Europan surface, with potential access to liquid water at depth, bolster the case for Europa's potential habitability and for future sampling of erupted material by spacecraft. Salinity and compositional gradients can affect the ice matrix, along with collapsing reservoirs, fracturing and diapirism, all of which may provide access to shallow water. Our results further expand the case that activity may be ongoing since the Galileo mission.

ACKNOWLEDGEMENTS

We thank A. Fruchter for helpful statistical discussions and E. Bergeron for assistance with the data. The data were obtained using the Hubble Space Telescope which is operated by STScI/AURA under grant NAS5-26555. We acknowledge support from grants associated with observing programs HST GO-13438, HST GO/DD-13620, HST GO-13829 and HST GO-14112. KPH acknowledge support from the Jet Propulsion Laboratory, California Institute of Technology, under a contract with the National Aeronautics and Space Administration.

REFERENCES

Allison, M.L. & Clifford, S.M. 1987, JGR, 92, 7865

Carlson, R.W., Calvin, W.M., Dalton, J.B. et al. 2009 in *Europa*, eds. R. Pappalardo, W. McKinnon, and K. Khurana, (University of Arizona Press), 283

Carr, M.H., Belton, M.J.S., & Chapman, C.R. et al. 1998, Nature, 391, 363

Chyba, C.E. 2000, Nature, 403, 381

Chyba, C.F. & Hand, K.P. 2001, Science, 292, 2026

Collins, G. & Nimmo, F. 2009 in *Europa*, eds. R. Pappalardo, W. McKinnon, and K. Khurana, (University of Arizona Press), 259

Craft, K.L., Patterson, G.W., Lowell, R.P. & Germanovich, L. 2016, Icarus, 274, 297

Crawford, G.D. & Stevenson, D.R. 1988, Icarus, 73, 66

Fagents, S.A., Greeley, R., Sullivan, R.J. et al. 2000, Icarus, 144, 54.

Fagents, S.A. 2003, JGR, 108, 5139 doi:10.1029/2003JE002128 (2003).

Figueredo, P.H., Greeley, R., Neuer, S., Irwin, L., & Schulze-Makuch, D. 2003, Astrobiology, 3, 851

Figueredo, P.H., Greeley, R. 2004, Icarus, 167, 287

Goldreich, P.M. & Mitchell, J.L. 2009, Icarus, 209, 631

Goodman, J.C., Collins, G.C., Marshall, J., & Pierrehumbert, R.T. 2004, JGR, 109, 3008 doi: 10.1029/2003JE002073 (2004).

Greenberg, R., Geissler, P., Hoppa, G., & Tufts, B.R. 2002, RvGeo, 40, 1004. doi:10.1029/2000RG000096

Hand, K.P., Carlson, R.W., & Chyba, C.F. 2007, Astrobiology, 7, 1006

Hand, K.P., Chyba, C.F., Priscu, J.C., Carlson, R.W., & Nealson, K.H. 2009 in *Europa*, eds. R. Pappalardo, W. McKinnon, and K. Khurana, (University of Arizona Press), 589

Hansen, C.J., Eposito, L., Stewart, A.I.F. et al. 2006, Science, 311, 1422

Hobbs, P.V. 1974, Ice Physics (Clarendon Press, Oxford) ISBN 0-19-851936-2

Howett, C.J.A., Spencer, J.R., Pearl, J., & Segura, M. 2010, Icarus, 206, 573

Kattenhorn, S.A., & Prockter, L.M. 2014, Nature Geoscience, 7, 762

Kivelson, M.G., Khurana, K.K., Russell, C.T., et al. 2000, Science, 289, 1340

Miyamoto, H., Mitri, G., Showman, A.P., & Dohm, J.M. 2005, Icarus, 177, 413

Nimmo, F., Spencer, J.R., Pappalardo, R.T., & Mullen, M.E., 2007, Nature, 447, 289

Ojakangas, G.W. & Stevenson, D.J. 1989, Icarus, 81, 242

Pappalardo, R.T., Belton, M.J.S., Breneman, H.H., et al. 1999, JGR, 104, 24015

Pappalardo, R.T., & Barr, A.C. 2004, GeoRL, 31, 1701

Phillips, C.B., McEwen, A.S., Hoppa, G.V. et al. 2000, JGR, 105, 22579

Phillips, C.B. 2015, 46th Lunar and Planetary Science Conference, LPI Contribution No. 1832, p.2704

Quick, L.C., Barnouin, O.S., Prockter, L.M., & Patterson, G.W. 2013, P&SS, 86, 1

Quick, L.C., Glaze, L.S., & Baloga, S.M. 2017, Icarus, 284, 477

Rathbun, J.A., Rodriguez, N.J., & Spencer, J.R. 2010, Icarus, 210, 763

Rhoden, A.R., Hurford, T.A., Roth, L., & Retherford, K. 2015, Icarus, 253, 169

Roth, L., Sauer, J., Retherford, K.D., et al. 2014, Science, 343, 171

Schmidt, B.E., Blankenship, D.D., Patterson, G.W., & Schenk, P.M. 2011, Nature, 479, 502

Sotin, C., Head, J.,W., Tobie, G. 2002, GeoRL, 29, 1233

Sparks, W.B., Hand, K.P., McGrath, M.A., et al. 2016, ApJ, 829, 121

Spencer, J.R., Tamppari, L.K., Martin, T.Z., & Travis, L.D. 1999, Science, 284, 1514

Walker, C.C., Bassis, J.N. & Liemohn, M.W. 2012, JGR, 117, E07003, doi:

10.1029/2012JE004084

Walker, C.C. & Schmidt, B.E. 2015, GeoRL, 42, 712

Zolotov, M.Yu. & Kargel, J.S. 2009 in *Europa*, eds. R. Pappalardo, W. McKinnon, and K. Khurana, (University of Arizona Press), 431

TABLE 1: OBSERVING PARAMETERS FOR 2016 OBSERVATIONS

| Rootname | Date | Exposure (seconds) | Filter (far ultraviolet, long-pass) | Longitude | True Anomaly | Angular Diameter (arcsec) |
|---|---|---|---|---|---|---|
| oczn05gcq | 2016-02-22 | 2192 | F25QTZ | 184.2 | 167 | 0.964 |
| oczn06d6q | 2016-03-04 | 2508 | F25QTZ | 180.5 | 169 | 0.971 |

FIGURE CAPTIONS

Figure 1. (a) Smoothed transit image from March 17, 2014, showing compact absorption feature near the limb at latitude ≈16.5°S, from Sparks et al. (2016) and discussed in depth there. (b) Smoothed transit image from February 22, 2016 showing a similar feature at the same location. (c) Average of images in (a) and (b) divided by the average of the other ten transits, showing absorption patch is transient, i.e. only present in these two images. (d) Image in (b) at full spatial resolution without smoothing, showing that the patch covers multiple pixels.

Figure 2. (a) The region near Pwyll, with the green ellipse outlining the position of the candidate of March 17, 2014, from Sparks et al. (2016). The transit viewing perspective is essentially tangential to the surface, from the right and the area covered by the image is ≈1320×900 km. The cyan ellipse is centered on the February 22, 2016 event, with dimensions approximating the uncertainty in its position. (b) Nighttime brightness temperature contours, in kelvins, from the Galileo PPR, from Spencer et al. (1999) showing a thermal anomaly with peak temperature at the same location as the plume candidates, within the uncertainties.

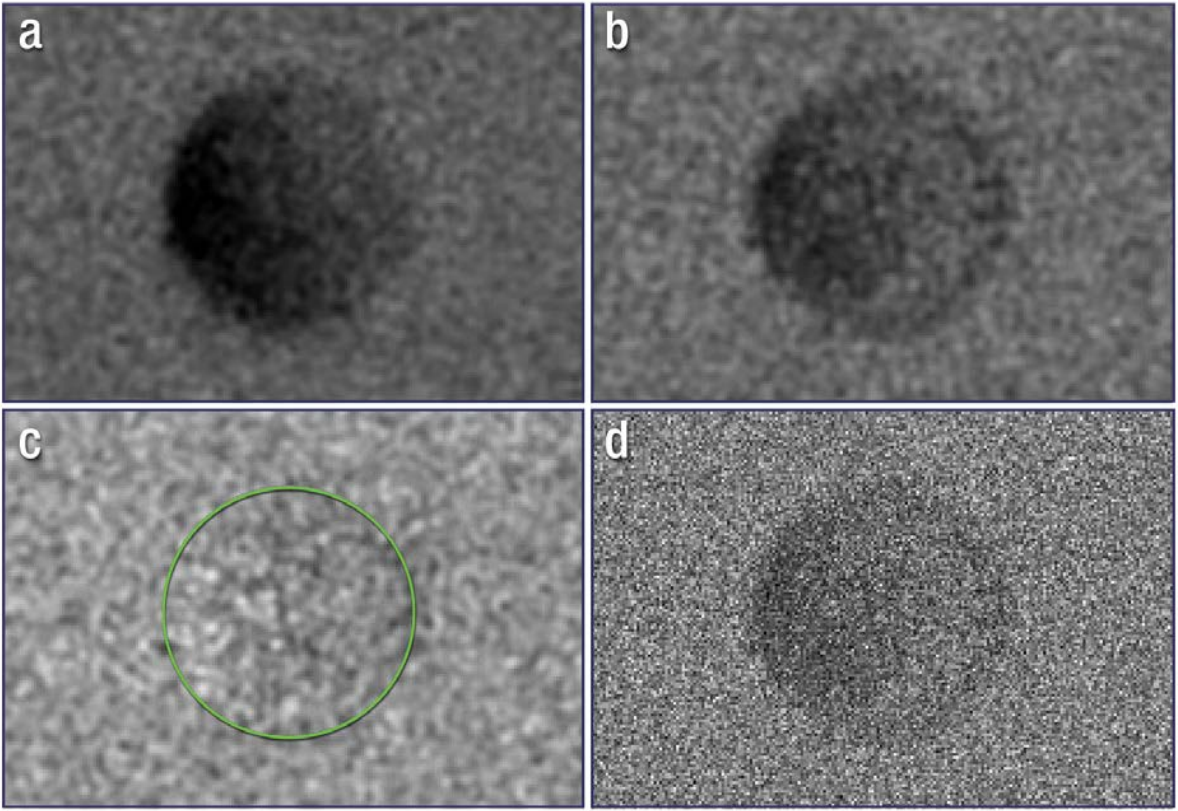

Figure 1

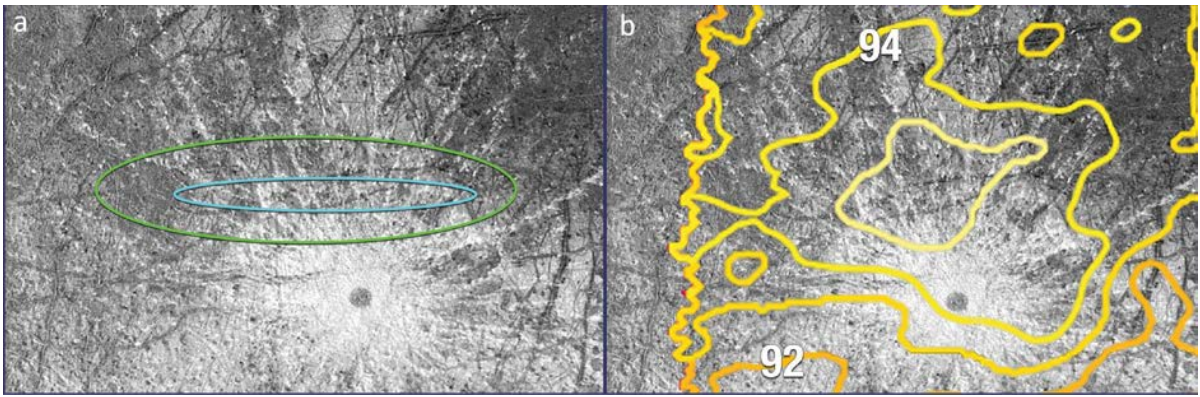

Figure 2